\documentclass[a4paper]{jpconf}
\usepackage{icsghead}

\usepackage[dvips]{graphicx}
\usepackage{amssymb,amsfonts,amsmath}
\usepackage{eucal,bm}
\usepackage{subfigure}
\usepackage{booktabs}
\usepackage{times,color,url}
\usepackage{tabularx}

\newcommand{\beq}{\begin{equation}}
\newcommand{\eeq}{\end{equation}}

\newcommand{\beqnr}{\begin{eqnarray}}
\newcommand{\eeqnr}{\end{eqnarray}}

\newcommand{\benum}{\begin{enumerate}}
\newcommand{\eenum}{\end{enumerate}}

\begin{document}

\title{Loop Calculus and Bootstrap-Belief Propagation\\ for Perfect Matchings on Arbitrary Graphs}

\author{Michael Chertkov $^a$, Andrew Gelfand $^{a,b}$, Jinwoo Shin $^c$}

\address{ Theoretical Divison \& Center for Nonlinear Studies,
Los Alamos National Laboratory,\\
Los Alamos, NM 87545, USA\\
Department of Computer Science,
University of California, Irvine,
Irvine, CA  92697-3435, USA\\
Mathematical Sciences Department,
IBM T. J. Watson Research,
Yorktown Heights, NY 10598, USA}

\ead{chertkov@lanl.gov}

\begin{abstract}
This manuscript discusses computation of the Partition Function (PF) and the Minimum Weight Perfect Matching (MWPM) on arbitrary, non-bipartite graphs. We present two novel problem formulations - one for computing the PF of a Perfect Matching (PM) and one for finding MWPMs - that build upon the inter-related Bethe Free Energy, Belief Propagation (BP), Loop Calculus (LC), Integer Linear Programming (ILP) and Linear Programming (LP) frameworks. First, we describe an extension of the LC framework to the PM problem. The resulting formulas, coined (fractional) Bootstrap-BP, express the PF of the original model via the BFE of an alternative PM problem.  We then study the zero-temperature version of this Bootstrap-BP formula for approximately solving the MWPM problem. We do so by leveraging the Bootstrap-BP formula to construct a sequence of MWPM problems, where each new problem in the sequence is formed by contracting odd-sized cycles (or blossoms) from the previous problem. This Bootstrap-and-Contract procedure converges reliably and generates an empirically tight upper bound for the MWPM.  We conclude by discussing the relationship between our iterative procedure and the famous Blossom Algorithm of Edmonds '65 and demonstrate the performance of the Bootstrap-and-Contract approach on a variety of weighted PM problems.
\end{abstract}

\section{Introduction}

Belief Propagation (BP) is an iterative message passing algorithm with physics roots in H. Bethe \cite{35Bet} and R. Peierls' \cite{36Pei} work on ``melting of lattices", information theory roots in decoding of Gallager's Low Density Parity Check (LDPC) codes \cite{63Gal}, and artificial intelligence roots dating to J. Pearl \cite{88Pea} (who actually coined the term). For decades the three fields developed BP and the homonymous Bethe-Peierls approximation independently. Only recently, under  the Bethe Free Energy (BFE) framework of Yedidia, Freeman and Weiss in \cite{01YFW,05YFW}, have the parallel developments been united.

BP is a naturally distributed and easy to implement algorithm for iteratively solving the system of BP equations - a set of equations that also characterize the fixed points of the BFE function~\cite{01YFW,05YFW}. The so-called BP/BFE approach can be used to compute the Partition Function (PF) or find the Maximum Likelihood (ML) assignment of a Graphical Model (GM), where a GM defines a joint probability over a set of random variables via local probability measures that factorize according to some underlying graph structure. BP/BFE provides an exact answer to the PF and ML tasks if the underlying graph is a tree. BP/BFE is a heuristic on graphs with loops, but it typically provides an accurate approximation.

Two recent developments have altered the paradigm that BP/BFE is only ``good'' for tree or tree-like GMs. First, it was shown that BP can exactly recover the ML assignment for certain binary GMs with dense, loopy graph structures. For example, BP can be used to find a PM on a bipartite graph \cite{05BSS,06BSS,08BSS}. Interestingly, all models of this special type can be formulated as Linear Programming (LP) problems that have an integral optimal solution, though this condition is not sufficient in general \cite{08BSS,08Che,11SMW}.

The Loop Calculus (LC) was developed at the same time and provides a way to express the PF a general binary GM as a finite (but still exponential in the size of the graph) series \cite{06CCa,06CCb}.  Each term in the series is related to a generalized loop of the underlying graph (which is a subgraph where all vertices have degree two or higher) and is expressed explicitly using the solution to the BP equations. In some special cases, e.g. for PMs over bipartite graphs where the partition function is a permanent \cite{10WC}, the multiplicative mismatch between the PF of the original problem and the respective BFE/BP expression is itself the PF of a new GM defined on the same graph. In such cases, the LC can be used to improve upon the BP estimate for the PF (when the BP estimate is not tight) by re-summing a number of important LC terms \cite{07GMK,10GKC}. As suggested in \cite{06CCc,11KJCa,11KJCb,13GCS}, the BP approximation to the ML problem can be improved by analyzing higher order terms of the LC (sometimes re-summing as in \cite{13GCS}) and using this information to modify the original GM. The hope is that when BP is run on the new GM it will be exact or, at least, improve upon its initial approximation.

In this manuscript we continue to work on improving BP, focusing on MWPM problems in general (non-bipartite) graphs.  This choice is not arbitrary.  MWPM is a special problem in Matching Theory which can be solved in polynomial time by Edmonds' famous Blossom Algorithm (BA)~\cite{Edmonds1965} (see~\cite{Kolmogorov2009} for details of BlossomV, the state-of-the-art implementation of the BA). The most recent development in this classical sub-field of Computer Science is due to Chandrasekaran, V{\'e}gh and Vempala \cite{12CVV} who proposed a polynomial time ``cutting-plane" procedure to finding a MWPM. In contrast to Edmonds' primal-dual algorithm, their approach carefully constructs a sequence of linear inequalities (or cut constraints) that tighten the LP relaxation, while ensuring that the added ``cuts'' maintain half-integrality of the LP.

While Edmonds' BA is efficient, it is not a distributed algorithm. Several recent results motivate the use of distributed, message passing implementations of BP to solve the MWPM problem. First, standard max-product BP was shown to be convergent and correct when the optimal solution to the LP relaxation of the weighted matching problem is integral. This was shown for PMs on bipartite graphs in~\cite{05BSS,06BSS,08BSS} and extended to general weighted matchings in~\cite{11SMW}. And in \cite{13GCS}, the MWPM was shown to be an example of a problem where the LP optima and iterative BP solution are equivalent (when the latter converges).

Unfortunately, the initial LP relaxation is not tight in general. The work in \cite{13GCS} considered situations when the optimal solution is non-integral and proposed running BP on a modified version of the weighted matching GM. In particular, higher order factors enforcing Edmonds' blossom (cut) constraints were added to the original GM and a finite temperature version of BP (with proper annealing and damping) was used to improve convergence to the LP optima. A provably accurate alternative to \cite{13GCS}, discussed in \cite{13CGSb}, would involve: (a) adaptively selecting the blossoms/contraints used to modify the GM; (b) maintaining half-integrality of the underlying LP at every iteration; and (c) proving convergence and correctness using the computational tree approach of \cite{08BSS,11SMW}.

This manuscript develops a new theoretical and algorithmic approach towards constructing efficient and distributed algorithms for resolving the MWPM problem. This new approach, which is complementary to both \cite{13GCS} and \cite{13CGSb}, is based on extending the LC paradigm of \cite{06CCa,06CCb,10WC,11YC} to PMs over an arbitrary graph at a finite temperature, $T$, and careful analysis of the ML, $T\to 0$, limit. Analysis of the $T\to 0$ limit led to the development of an algorithm for (approximately) solving the MWPM problem - the BP-Bootstraph-Contract procedure - which we evaluated empirically.

The material in the manuscript is organized as follows. We provide some background on Graphical Models (GMs) and Perfect Matching (PM) problems in Section \ref{sec:GM}; the BP/BFE framework is briefly reviewed in Section \ref{sec:BP} and in \ref{app:conv}. The main results involving the LC are covered in Section \ref{sec:LS}. Finally, the Bootstrap BP and fractional Bootstrap BP (\cite{11YC}) formulas are discussed in Section \ref{sec:boot} and \ref{app:boot}. The ML, $T\to 0$, versions of the Bootstrap formulas are discussed and illustrated on a small example in Section \ref{sec:T0}. The Bootstrap-and-Contract algorithm is presented in Section \ref{sec:algorithm} and evaluated empirically in Section \ref{sec:experiments}. Finally, we summarize and discuss open questions in Section \ref{sec:path}.

\section{Graphical Model of Perfect Matching}
\label{sec:GM}

Consider an arbitrary undirected graph, ${\cal G}=({\cal V},{\cal E})$, with associated edge weights
$w=(\exp(-\varepsilon_{ij}/T)|(i,j)\in{\cal E})$, where $T$ is a positive temperature parameter term and $\varepsilon_{ij}$ is the energy of a Perfect Matching (PM), or dimer\footnote{The two terms, used respectively in Statistical Physics and Computer Science, are equivalent.}, associated with the edge $(i,j)$.  One problem of interest is to evaluate the PF of the PM over the tuple $({\cal V},{\cal E},w)$:
\begin{equation}
Z(w)=\sum_{\sigma\in\Sigma} \exp\left(-\frac{1}{T}\sum_{(i,j)\in{\cal E}}\varepsilon_{ij}\sigma_{ij}\right)=\sum_{\sigma\in\Sigma} (w_{ij})^{\sigma_{ij}}=
\sum_{\sigma\in\Sigma} W(\sigma),
\label{Z}
\end{equation}
where  $\Sigma= \left \{ \sigma|\forall (i,j): \sigma_{ij}=\{0,1\}; \;
\forall i: \; \sum_{j:(i,j)\in {\cal E}}\sigma_{ij}=1 \right \}$ is the set of PMs in ${\cal G}$, where a PM is a subset of the edges in ${\cal E}$ such that an edge is adjacent to each vertex in ${\cal V}$. Note that ${\cal P}(\sigma)=W(\sigma)/Z(w)$ is thus the probability of the PM $\sigma$.

The total weight, $E$, of the Maximum Weight Perfect Matching (MWPM) arises by taking the zero temperature limit, $T\to 0$, of the PF
\begin{equation}
E=-\lim_{T\to 0} \frac{\log Z}{T}=\min_{\sigma\in\Sigma}\sum_{(i,j)\in{\cal E}}\varepsilon_{ij}\sigma_{ij}.
\label{E}
\end{equation}
The rhs of Eq.~(\ref{E}) is an Integer Linear Programming (ILP) problem. The LP relaxation of Eq.~(\ref{E}) is
\begin{equation}
E_0=\min_{x\in X}\sum_{(i,j)\in{\cal E}}\varepsilon_{ij}x_{ij},
\label{E0}
\end{equation}
where $X=\left \{ x_{ij}|\forall (i,j): x_{ij}\in[0,1]; \; \forall i: \; \sum_{j:(i,j)\in {\cal E}}x_{ij}= 1 \right \}$ is the so-called PM polytope. Since Eq.~(\ref{E0}) constitutes a relaxation of Eq.~(\ref{E}), $E_0\leq E$.  As shown in \cite{03Sch} the optimal $x$ in Eq.~(\ref{E0}) is half integer.

If the graph is bipartite then (a) computing the PF in (\ref{Z}) is equivalent to computing the permanent of the matrix of $w$-weights; (b) the PM polytope, $X$ is equal to the convex hull of $\Sigma$; (c) $E_0=E$,  i.e. the LP-relaxation is gapless; and (d) the optimal $x$ in Eq.~(\ref{E0}) is integral (assuming the formulation in (\ref{E}) is not degenerate).

\section{Bethe Free Energy, Belief Propagation and LP-BP}
\label{sec:BP}

For a generic Graphical Model (GM) over ${\cal G}$ assigning (un-normalized) weight $w$ to state $\sigma$, one defines the exact variational~\footnote{It is referred to as the Gibbs and Kullback-Leibler function in statistical physics and statistics, respectively.} function
\begin{eqnarray}
{\cal F}\{b\}\equiv T \sum_{ \sigma }b(\sigma)\ln\frac{b(\sigma)}{W(\sigma)}. \label{Gibbs}
\end{eqnarray}
The belief, $b(\sigma)$ is understood as a proxy to the probability ${\cal P}(\sigma)$ because under the normalization condition $\sum_{\sigma \in \Sigma} b(\sigma)=1$, the Gibbs function is convex and it achieves its only minimum at $b(\sigma)={\cal P}(\sigma)$ and ${\cal F}\{ {\cal P} \}=-T\ln Z$.

Belief Propagation (BP) is an iterative method for computing beliefs that is exact when the underlying GM is a tree.
As shown in \cite{05YFW}, the BP fixed point equations can be derived as a relaxation of the constrained Gibbs function (\ref{Gibbs}). We briefly review the concepts of \cite{05YFW} with application to the PM problem. The material generalizes the description of \cite{10WC}, which was limited to bipartite graphs (permanents).

The BP approximate belief of state $\sigma\in\Sigma$ is
\begin{eqnarray}
 b(\sigma)\approx b_{\it BP}(\sigma)=
 \frac{\prod_i b_i(\sigma_i)}{
 \prod_{(i,j) \in {\cal E}} b_{ij}(\sigma_{ij})},
\label{BP_Belief}
\end{eqnarray}
where $b_i(\sigma_i)$ ($b_{ij}(\sigma_{ij})$) are vertex (edge) beliefs related to each other by
\begin{equation}
\forall (i,j) \in {\cal E} :\quad b_{ij}(\sigma_{ij})= \sum\limits_{\sigma_i\setminus\sigma_{ij}}b_i(\sigma_i)=
\sum\limits_{\sigma_j\setminus\sigma_{ij}}b_j(\sigma_j),\label{rel}
\end{equation}
and where $\sigma_i=(\sigma_{ij}|j: (i,j)\in{\cal E})$ is a vector of edge beliefs.
The beliefs should also satisfy the normalization conditions:
\begin{equation}
\forall (i,j) \in E :\quad b_{ij}(1)+b_{ij}(0)=1.\label{norm}
\end{equation}
Substituting Eq.~(\ref{BP_Belief}) into Eq.~(\ref{Gibbs}) and utilizing Eqs.~(\ref{rel},\ref{norm}) one arrives at the following Bethe Free Energy function (BFE)
\begin{eqnarray}
&&  {\cal F}_{\mbox{BP}}(\beta)\equiv E-T S, \quad E\equiv\sum_{(i,j)}b_{ij}(1)\varepsilon_{ij},\label{FE}\\
&& S\equiv\sum_{(i,j)\in{\cal E}}\sum\limits_{\sigma_{ij}}
  b_{ij}(\sigma_{ij})\ln b_{ij}(\sigma_{ij})
   -\sum_{i\in{\cal V}}\sum\limits_{\sigma_i} b_i(\sigma_i).\label{S}
\end{eqnarray}

Eqs.~(\ref{rel},\ref{norm},\ref{FE},\ref{S}) can be greatly simplified. One can express Eqs.~(\ref{rel},\ref{norm}) solely in terms of $\beta_{ij}\equiv b_{ij}(1)$ satisfying the following Doubly Stochastic (DS) constraints
\begin{eqnarray}
 \forall (i,j) \in {\cal E} : 0\leq \beta_{ij}\leq 1; \quad
\forall i:  \sum_j \beta_{ij}=1. \label{ds_cond}
\end{eqnarray}

We will use $\beta\in \mbox{DS}$ to indicate that the matrix $\beta$ is DS,  i.e. lies in the $\mbox{DS}$ polytope defined by Eq.~(\ref{ds_cond}).
The entropy Eq.~(\ref{S}) becomes
\begin{equation}
S(\beta)=\sum_{(i,j)\in{\cal E}}\left((1-\beta_{ij})\ln(1-\beta_{ij})-\beta_{ij}\ln\beta_{ij}\right). \label{ES}
\end{equation}

Therefore, the BFE/BP approximation to the partition function of the PM model over ${\cal G}$ becomes
\begin{eqnarray}
&& -\log Z_{\mbox{BP}}= \frac{1}{T}\min_{\beta\in \mbox{DS}} {\cal F}_{\mbox{BP}}(\beta)\nonumber\\
&&=\min_{\beta\in \mbox{DS}} \sum_{(i,j)\in{\cal E}}\left(\beta_{ij}\left(\frac{\varepsilon_{ij}}{T}+\log \beta_{ij}\right)-(1-\beta_{ij})\log(1-\beta_{ij})\right).
\label{BFE}
\end{eqnarray}

The MWPM can also be estimated within the BFE/BP approach as:
\begin{equation}
E_{\mbox{BP}}=\min_{\beta\in\mbox{DS}} \beta_{ij}\varepsilon_{ij}.
\label{E_BP}
\end{equation}
Notice that Eq.~(\ref{E_BP}) is an LP, that will be called BPLP because of its relation to the BP concept.
As shown in \cite{08BSS}, $E_{\mbox{BP}}=E_0$, i.e. LP$\equiv$BPLP, where LP is defined by Eq.~(\ref{E0}). One also finds that the optimal $\beta$ in Eq.~(\ref{E_BP}) is half-integral.

For a most general GM the BFE function is bounded from below but non-convex.  However, as shown in \cite{13Von} for the PM over bi-partite graph,  ${\cal F}_{\mbox{BP}}(\beta)$ is a convex function over $\beta\in\mbox{DS}$. As argued in \ref{app:conv} the BFE convexity proof of \cite{13Von} extends straightforwardly to PM GM over arbitrary ${\cal G}$.

To analyze the minima of the constrained BFE in Eq.~(\ref{BFE}) we incorporate Lagrange multipliers $\mu_i$ enforcing the constraints in Eqs.~(\ref{ds_cond}). Looking for a stationary point of the respective Lagrange function over the $\beta$ variables, one arrives at the following set of quadratic equations for edge variables, $\beta_{ij}$
\begin{equation}
 \forall (i,j) \in E :\quad
 \beta_{ij}(1-\beta_{ij})=\exp\left(\frac{\mu_i+\mu_j-\varepsilon_{ij}}{T}\right). \label{BP1}
\end{equation}
Eqs.~(\ref{BP1}) combined with the conditions (\ref{ds_cond}) constitute the so-called Belief Propagation (BP) equations.

To find a solution of BP Eqs.~(\ref{ds_cond},\ref{BP1}) one relies on an iterative procedure. For a description of a set of iterative BP algorithms convergent to a minimum  of the BFE for the perfect matching problem we
refer the interested reader to \cite{08CKV,10CKKVZ,09HJ}.

\section{Loop Series}
\label{sec:LS}

As shown in \cite{06CCa,06CCb}, the exact partition function of a generic GM can be expressed in
terms of a Loop Series (LS), where each term is computed explicitly using beliefs found by solving the BP equations.

Adapting this general result to the PM problem one derives
\begin{eqnarray}
 && Z=Z_{\mbox{BP}}\cdot z,\quad z\equiv 1+\sum_{C} r_{C},\quad
 r_{C}=\left(\prod_{i\in{C}}\psi_{i;{C}}\right),\label{Loop}\\
 && \psi_{i;{\it
 C}}\equiv\frac{\sum_{\vec{\sigma}_i}b_i(\sigma_i)
 \prod_{j:(i,j)\in{C}}\left(m_{ij}-2\sigma_{ij}+1\right)}
 {\prod_j^{(i,j)\in{C}}\sqrt{1-(m_{ij})^2}}
 =(1-q_i)\prod_{j\in{C}:(i,j)\in{C}}
 \sqrt{\frac{\beta_{ij}}{1-\beta_{ij}}}
 ,\label{psi_i}\\ &&
 m_{ij}\equiv\sum_{\sigma_{ij}}(2\sigma_{ij}-1)b_{ij}(\sigma_{ij})=2\beta_{ij}-1,\quad
 q_i\equiv\sum_{j\in{C}}^{(i,j)\in{C}}1,\label{mq}
\end{eqnarray}
where once again $\{b\}$ (or $\{\beta\}$) are solutions to the BP-equations Eqs.~(\ref{ds_cond},\ref{BP1}) and $C$ stands for an arbitrary generalized loop, defined as a subgraph of the complete bipartite graph with all its vertices having a degree larger than 1. The $q_i$ in Eq.~(\ref{rC}) are the $C$-dependent degrees, i.e. $q_i=\sum_{j \mid (i,j)\in C} 1$.

Let us clarify derivation of the compact expressions on the rhs of
Eqs.~(\ref{psi_i}):
\begin{eqnarray}
 && \sum_{\sigma_i}b_i(\sigma_i)
 \prod_{j:(i,j)\in{C}}\left(m_{ij}-2\sigma_{ij}+1\right)=
 \sum_{k:(i,k)\notin{C}} \beta_{ik}\prod_{j:(i,j)\in{C}} (m_{ij}+1)\nonumber\\ &&
 +\sum_{k:(i,k)\in{C}} \beta_{ik}(m_{ik}-1)\prod_{j\neq k}^{(i,j)\in{\it
 C}}(m_{ij}+1)=2^{q_i}(1-q_i)\prod_{j:(i,j)\in{C}}\beta_{ij},\label{der}
\end{eqnarray}
where we used relations (\ref{mq}) and also the double stochasticity of $\beta$.

Combining Eqs.~(\ref{Loop},\ref{psi_i}) one derives
\begin{eqnarray}
 r_c=\left(\prod_{i\in {C}} (1-q_i)\right)\prod_{(i,j)\in{C}}
  \frac{\beta_{ij}}{1-\beta_{ij}}.
 \label{rC}
\end{eqnarray}
In Eq.~(\ref{rC}), loops comprised of an even (odd) number of vertices give positive (negative) contributions to $r_C$. Eq.~(\ref{rC}) was stated in \cite{10WC} for the bipartite case. From the above derivation, we can see that the result is valid for any graph ${\cal G}$.

\section{Bootstrapping BP and Fractional BP}
\label{sec:boot}

Another important result for PMs is the following re-summation of the LS
\begin{equation}
 \frac{Z(w)}{Z_{\mbox{BP}}(w)}= \frac{Z(\tilde{\beta})}{ \prod_{(i,j)\in {\cal E}}(1-\beta_{ij})}, \label{tilde_beta}
\end{equation}
where $\tilde{\beta}=(\beta_{ij}(1-\beta_{ij})|(i,j)\in{\cal E})$. Eq.~(\ref{tilde_beta}) was stated in \cite{10WC} only for the case of permanent - PM over binary graph,  even though direct verification confirms (see \ref{app:boot}) that the formula is generic, applicable to PM over an arbitrary graph.

We call Eq.~(\ref{tilde_beta}) the Bootstrap-BP equation because the rhs of Eq.~(\ref{tilde_beta}) provides a quantitative measure of the relative accuracy of BP.

Extending the approach of \cite{11CY} from bipartite to general graphs, one arrives at the following fractional generalization of Eq.~(\ref{tilde_beta})
\begin{eqnarray}
 && \frac{Z(w)}{Z_\gamma(w)}= Z(\tilde{\beta}_\gamma) \prod_{(i,j)\in {\cal E}}(1-\beta_{ij})^{\gamma}, \label{tilde_beta_gamma}\\
 && -\log Z_\gamma(w)= \min_{\beta\in \mbox{DS}} \sum_{(i,j)\in{\cal E}}\left(\beta_{ij}\left(\frac{\varepsilon_{ij}}{T}+\log \beta_{ij}\right)+\gamma(1-\beta_{ij})\log(1-\beta_{ij})\right),\label{Zgamma}
\end{eqnarray}
where $\tilde{\beta}_\gamma=(\beta_{ij}*(1-\beta_{ij})^{-\gamma}|(i,j)\in{\cal E})$ and the doubly stochastic
$\beta_{ij}$'s entering Eqs.~(\ref{tilde_beta},\ref{Zgamma}) come from solving the following $\gamma$-version of Eq.~(\ref{BP1})
\begin{equation}
 \forall (i,j) \in {\cal E} :\quad
 \beta_{ij}(1-\beta_{ij})^{-\gamma}=\exp\left(\frac{\mu_i+\mu_j-\varepsilon_{ij}}{T}\right), \label{BP-gamma}
\end{equation}
$\gamma$ can be any number in $[-1;1]$ ,  with $\gamma=-1$ corresponding to the BP case.
Notice, that the Loop Series representation for the partition function correction only applies to the BP case.
Its generalization to arbitrary $\gamma$ is not known to the authors.

\section{Finding MWPMs via BP: An Illustration}
\label{sec:T0}

For an arbitrary graph, the BPLP=LP relaxation of the MWPM problem is known to be half-integral \cite{11SMW}. When the solution to the relaxation is integral, one has a certificate of optimality and $E=E_0$. When the solution is half integral, $E_0<E$, simply because BPLP is a relaxation of the original MWPM problem. The Bootstrap BP recasting of the LS provides a compact expression relating $E$ and $E_0$ in terms of some new partition function. The question explored in the remainder of this section is the following:
\begin{itemize}
\item Can the (fractional) Bootstrap-BP formula in Eq.~(\ref{tilde_beta_gamma}) be used to close the gap between $E_0$ and $E$?
\end{itemize}

The rhs of Eq.~(\ref{tilde_beta_gamma}) involves the PF of a new GM defined on graph ${\cal G}$. Our hope is that the new GM is easier to deal with than the original GM. Before checking to see if this is true, we first analyze the $T\to 0$ limit of Eq.~(\ref{tilde_beta_gamma}).

The fractional parameter $\gamma$ in Eq.~(\ref{tilde_beta_gamma}) can be chosen arbitrary from the interval $[-1;1]$. However, in order to avoid resolving a fictitious singularity~\footnote{
The algebraic factor on the rhs of Eq.~(\ref{tilde_beta_gamma}) is singular in the $T\to 0$ limit at $\gamma<-1$ when some $\beta_{ij}$ approaches unity. However, the singularity is fictitious, as the PF on the rhs of Eq.~(\ref{tilde_beta_gamma}) will turn to $0$ (in the limit). Since
$Z(w)\sim \exp(-E/T)$ and $Z_\gamma(w)\sim \exp(-E_0/T)$ and $E_0<E$ in the case of interest, we expect to get zero (in the limit) for the rhs of Eq.~(\ref{tilde_beta_gamma}).} it is convenient to consider $\gamma=0$. Then in the $T\to 0$ limit one derives from Eq.~(\ref{tilde_beta_gamma},\ref{BP-gamma}):
\begin{eqnarray}
E=E_0+E_1, \quad E_1=\min_{\sigma\in \Sigma}\sum_{(i,j)\in{\cal E}}(\varepsilon_{ij}-\mu_i-\mu_j)\sigma_{ij}
\label{E-E0}
\end{eqnarray}
where $E$ and $E_0$ are as in Eqs.~(\ref{E},\ref{E0}), and $\mu$ are dual variables (often called chemical potentials in statistical physics) that can be reconstructed from the optimal solution of the BPLP (\ref{E0}) as follows:
\begin{itemize}
\item
One first identifies any non-intersecting half-integral cycles $C$ of ${\cal G}$, i.e. odd-length cycles where  $\beta_{ij}=1/2$ for each edge $(i,j) \in C$  and $\{\beta_{ij}\}$ is the solution of BPLP~\footnote{If the BPLP is fractional, then at least two such cycles exist. This is guaranteed by half-integrality of BPLP and because we are solving a PM problem.}.

\item One computes Lagrangian multipliers, $\mu_i$, along each half-integral cycle, resolving, separately for every cycle the set of linear equations, $\mu_i+\mu_j=\varepsilon_{ij}$.

\item For every edge, $(i,j)$ with $\beta_{ij}=1$ one computes the Lagrangian multipliers associated with the vertices $i$ and $j$, $\mu_i\leftarrow \varepsilon_{ij}/2+\delta_{ij}$ and $\mu_j\leftarrow\varepsilon_{ij}/2-\delta_{ij}$, where $\delta_{ij}$ can be chosen arbitrarily,  for example $\delta_{ij}\leftarrow\varepsilon_{ij}/2$.~\footnote{The choice of $\delta$ does not affect the optimal $\sigma$ defining $E_1$ in Eq.~(\ref{E-E0}).}
\end{itemize}

\begin{figure}
\centering
\subfigure[]{\includegraphics[width=0.45\textwidth,page=2]{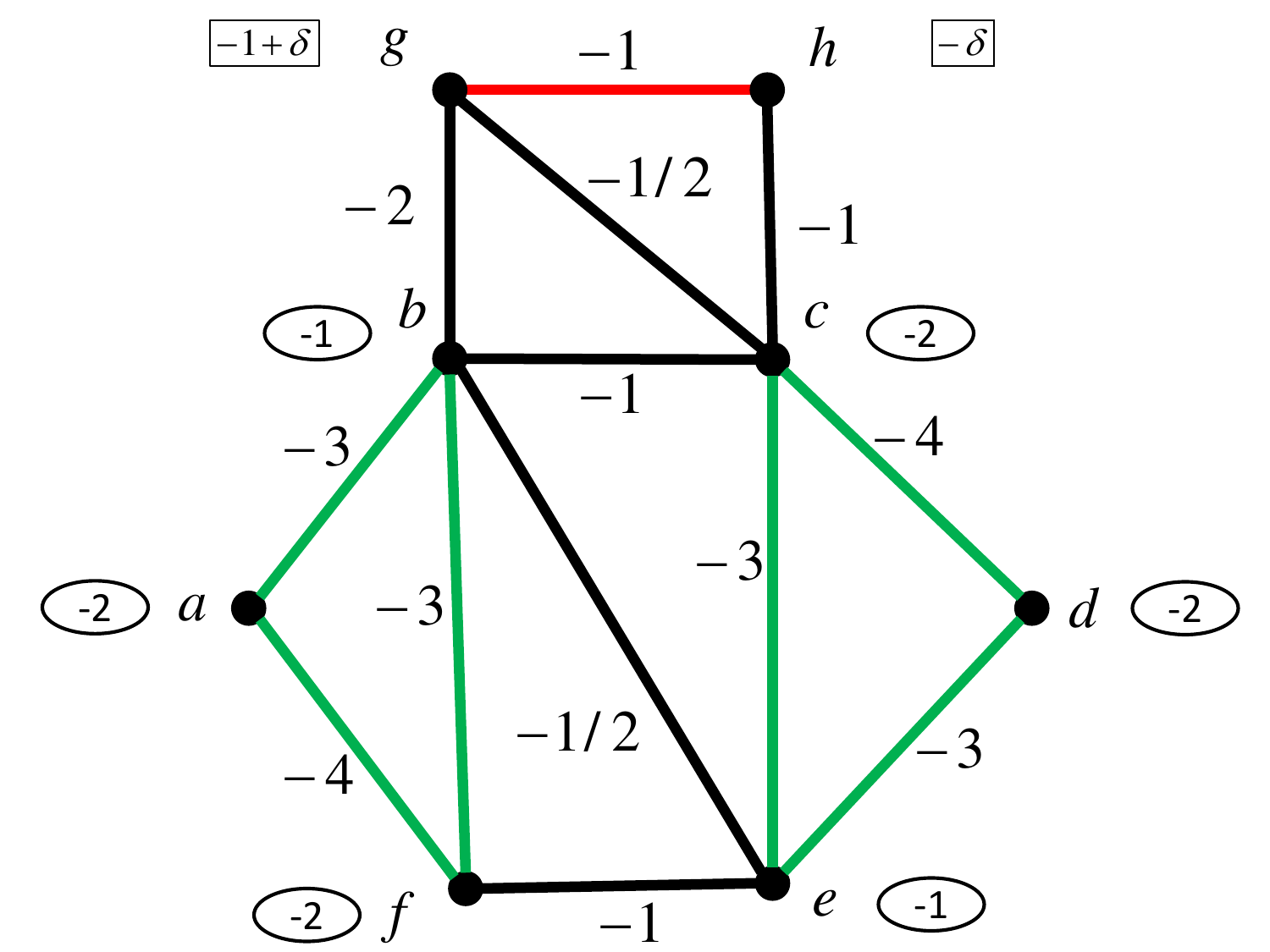}}
\subfigure[]{\includegraphics[width=0.45\textwidth,page=3]{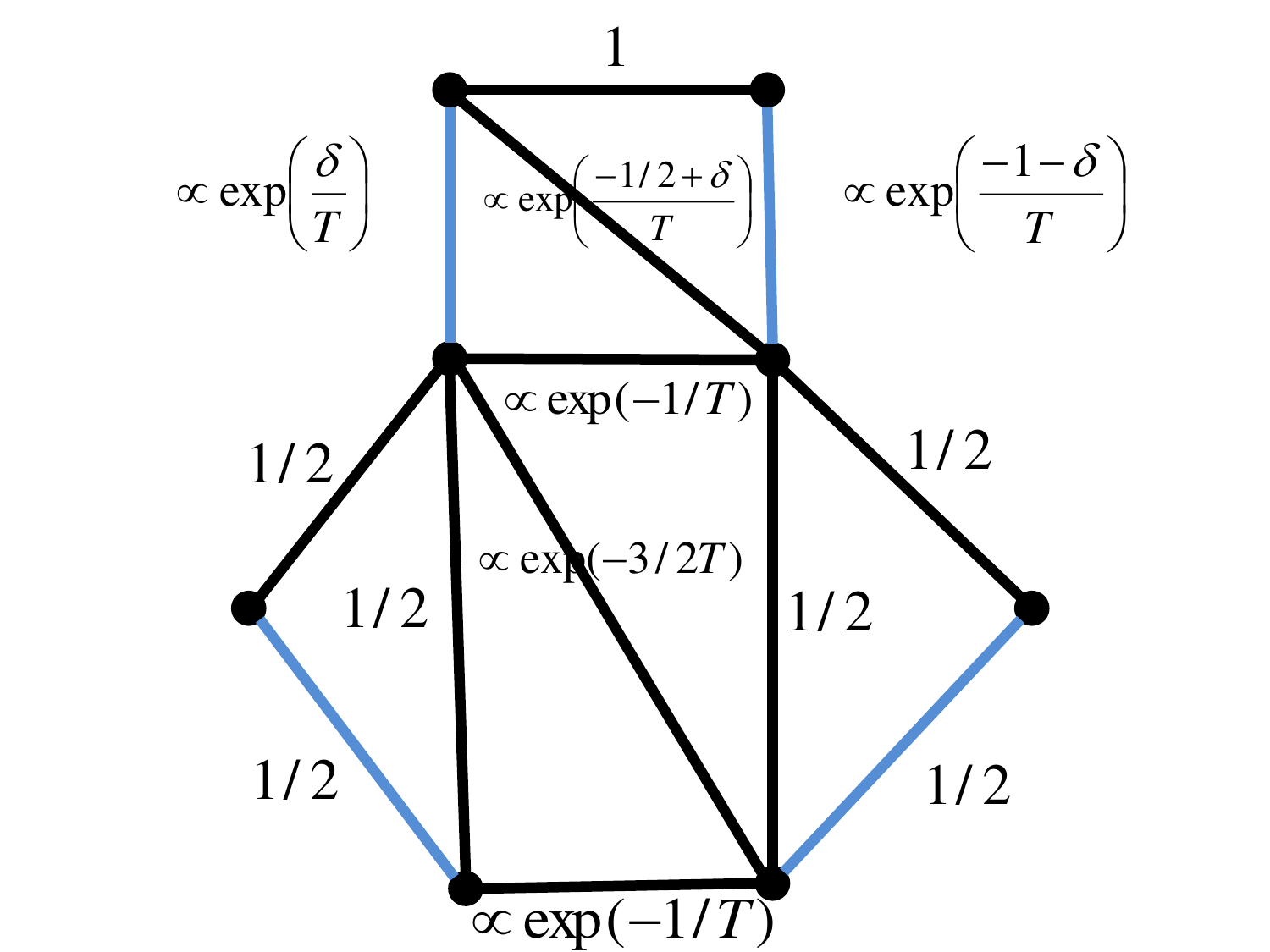}}
\caption{Illustrative $8$-node example analysed with the $\gamma=0$ version of the bootstrap-BP approach. (a) Numbers next to edges are $\varepsilon_{ij}$. Edges shown in green and red output $\beta_{ij}=1/2$ and $\beta_{ij}=1$ respectively in BP-LP. MWPM is $(a,f),(e,d),(g,b),(h,c)$. Numbers shown in ovals and squares correspond to the leading asymptotic estimates for the Lagrangian multipliers, $\mu_i$. Freedom in choosing the parameter $\delta$, entering $\mu_i$ adjusted to $\beta_{ij}=1$ does not affect the $E-E_0$ correction we are looking for in here. (b) Expressions next to the edges are respective $\beta_{ij}$.  Edges marked blue give the leading order PM contribution to $Z(\tilde{\beta})$.}
\label{fig:8nodes1}
\end{figure}

To demonstrate utility of the bootstrap BPLP formulas (\ref{E0},\ref{E-E0}) let us consider the $8$-node
example shown in Fig.~(\ref{fig:8nodes1}). In this example, the MWPM is $(a,f),(e,d),(g,b),(h,c)$ and, therefore,
$E=-4-3-2-1=-10$. The solution to the BPLP is fractional. All edges on the two cycles, $(a,b,f)$ and $(c,d,e)$ have $\beta=1/2$, edge $(g,h)$ has $\beta_{gh}=1$ and all other edges in the graph have $\beta=0$.  Therefore,  $E_0=(-4-3-3)/2+(-4-3-3)/2-1=-11$, which is clearly less than $E$.  All dual variables, $\mu_i$, associated with the half-integral cycles (shown in green in Fig.~(\ref{fig:8nodes1}a)) are determined unambiguously. The dual variables for edge $(g,h)$ must satisfy $\mu_g+\mu_h+1=0$ and we choose $\mu_g-\mu_h=2\delta$. In principle, the freedom can be fixed taking into account the DB vertex constraints on all the $\beta$-corrections. However, this is not necessary because the $\delta$'s simply cancel when computing the correction $E_1=\delta+1-\delta=1=E_0-E$ for the PM contribution to $Z(\beta)$ marked in blue in Fig.~(\ref{fig:8nodes1}b).

A number of things are worth noting from this simple example. First, the Bootstrap formula (\ref{tilde_beta})
allowed us to reduce the original MWPM problem to an auxiliary MWPM problem on the same graph. At first,
the new MWPM problem does not appear to be easier to solve than the original MWPM problem. However, this
example suggests a direction to further simplify the auxiliary problem by transforming it to a MWPM problem
on a strictly smaller graph. Indeed, for the example considered one can \emph{contract} each of the ``green, half-integer" cycles to individual nodes. This suggestion is inspired by the following observations~\cite{Edmonds1965,Kolmogorov2009,12CVV}:
\begin{itemize}
\item Define a \emph{blossom} to be an odd-sized cycle and choose our blossoms to be the odd-sized cycles with $\beta_{ij}=1/2$ -- marked green in Fig.~(\ref{fig:8nodes1}a). Such blossoms contribute zero weight to the energy correction $E_1$.

\item Edmond's ``outer" constraint associated with any of the ``half-integer" (green) blossom states that {\it at least one edge} from the cut (outer) edges should be in the optimal MWPM solution of the original problem. Note that for any of the ``half-integer" blossoms in the example of Fig.~(\ref{fig:8nodes1}) exactly one cut edge (and not more than one!) contributes to the optimal MWPM of the original problem.
\end{itemize}

In the following section we describe an algorithm that repeatedly applies this contraction concept to identify
a sequence of corrections to the MWPM energy $E$.

%Our numerical tests (discussed below) suggest that contraction of the ``half-integer" blossoms fails on a small number of dense graphs instances. This observation justifies using the contraction step empirically in the provably convergent algorithm for finding MWPM discussed in the next Section.

\section{Algorithm for finding MWPM sequentially}
\label{sec:algorithm}

\begin{center}
  \fbox{
       \begin{minipage}{0.95\linewidth}
       \begin{center}
       %\begin{ALG}{MWPM Sequential Reduction}\label{alg}\end{ALG}
       {\bf \underline{Algorithm:} BP-Bootstrap-Contract-Repeat}
       \end{center}
       {\bf Initialization:} The master system is defined as a non-degenerate MWPM over the $({\cal V},{\cal E},w)$ tuple. (Add small random corrections to $w$ to ensure non-degneracy.). Set $E=0$.\\ \\
       {\bf Iterate:} \\ \\
       {\bf (1)} Solve the LP=BP-LP, Eqs.~(\ref{E0},\ref{E_BP}) over $({\cal V},{\cal E},w)$.
       (For efficient distributed implementation use the algorithm of \cite{08BSS,11SMW}.)
       $\beta$ and $E_0$ are the outputs. Set, $E+=E_0$.
       \\  \\
       {\bf (2)} If solution of the LP=BP-LP is integer: {\bf Stop}.\\ \\
       {\bf (3)} Otherwise, modify the master system following the procedure:\\
       \begin{itemize}
       \item {\bf (a)} Identify non-intersecting half-integral cycles of ${\cal G}$, i.e. cycles along which all edged are half-integral, $\beta_{ij}=1/2$, in the solution of the LP=BP-LP. (Existence of at least two such cycles is guaranteed.)

       \item {\bf (b)} Compute the Lagrangian multipliers, $\mu_i$, along the half-integer cycles,  resolving, separately for every cycle the set of linear equations, $\mu_i+\mu_j=\varepsilon_{ij}$.

       \item {\bf (c)} For every edge, $(i,j)$, which is in PM according to the LP=BP-LP, i.e. $\beta_{ij}=1$, compute the Lagrangian multipliers associated with the vertices $i$ and $j$, $\mu_i\leftarrow \varepsilon_{ij}/2+\delta_{ij}$ and $\mu_j\leftarrow\varepsilon_{ij}/2-\delta_{ij}$, where $\delta_{ij}$ can be chosen arbitrarily,  for example $\delta_{ij}\leftarrow\varepsilon_{ij}/2$.

       \item {\bf (d)} Update weights, $\forall (i,j):\ \varepsilon_{ij}\leftarrow\varepsilon_{ij}-\mu_i-\mu_j$.
       %w_{ij}=\exp((\mu_i+\mu_j-\varepsilon_{ij})/T)$.

       \item {\bf (e)} Update the $({\cal V},{\cal E},w)$ tuple according to the following rules: Contract all vertices correspondent to a half-integer cycle into one new vertex. If multiple edges have appeared in the result of the contraction, combine them in one edge choosing the new weight equal to be the largest weight of the combined edges. Repeat it for all the half-integer cycles.

       \item {\bf (f)} Add small random correction to $w$ to remove possible (future) degeneracy.

       \end{itemize}
\vspace{2pt}
\end{minipage}
      }
\vskip 3pt
\end{center}

As already noted, the BP-Bootstrap procedure outputs an energy $E$ which is an upper bound on the value of the MWPM (proof of this statement will be presented in a future work). Once the upper bound is derived, one can use the following heuristics to find the PM corresponding to the output. Assume that all the $\epsilon$ weights in the original PM formulation are integer and solve the following auxiliary LP,
$$ min \sum_{(i,j)\in{\cal E}} \varepsilon_{ij} x_{ij},\ \mbox{s.t. } \sum_{(i,j)\in{\cal E}} \varepsilon_{ij} x_{ij}\leq E,\ \forall i\in{\cal V}:\ \sum_{j:(i,j)\in{\cal E}} x_{ij} = 1,\quad \forall (i,j)\in{\cal E}\ x_{ij}\geq 0.$$
If the LP outputs an integral solution, we stop. Otherwise, we run the same LP with $E-1$ instead of $E$, and repeat the procedure till an integral solution is found. 

\section{Experiments}
\label{sec:experiments}

\begin{table}
\begin{center}
\begin{tabularx}{0.745\textwidth}{|l||c|c|c||l||c|c|c|} \hline
  \multicolumn{4}{|c||}{50 \% sparse} & \multicolumn{4}{c|}{90 \% sparse} \\
\scriptsize{N / M} & \scriptsize{\# Cor.} & \scriptsize{\# LPs} & \scriptsize{\# Blos.} &
\scriptsize{N / M} & \scriptsize{\# Cor.} & \scriptsize{\# LPs} & \scriptsize{\# Blos.} \\ \hline

\scriptsize{100 / 1961} & \scriptsize{100} & \scriptsize{1.4[1,4]} & \scriptsize{0.9[0,6]} &
\scriptsize{100 /475} & \scriptsize{100} & \scriptsize{1.4[1,4]} & \scriptsize{0.8[0,8]} \\ \hline

\scriptsize{250 / 12284} & \scriptsize{100} & \scriptsize{1.6[1,5]} & \scriptsize{1.2[0,8]} &
\scriptsize{250 / 2973} & \scriptsize{100} & \scriptsize{1.6[1,3]} & \scriptsize{1.2[0,4]} \\ \hline

\scriptsize{500 / 49159} & \scriptsize{100} & \scriptsize{1.8[1,6]} & \scriptsize{1.7[0,12]} &
\scriptsize{500 / 11898} & \scriptsize{100} & \scriptsize{1.7[1,4]} & \scriptsize{1.5[0,6]} \\ \hline

\scriptsize{1000 / 196696} & \scriptsize{100} & \scriptsize{1.9[1,5]} & \scriptsize{1.9[0,8]} &
\scriptsize{1000 / 47576} & \scriptsize{99} & \scriptsize{1.8[1,4]} & \scriptsize{1.8[0,8]} \\ \hline

\end{tabularx}
\caption{Evaluation of BP-Bootstrap-Contract on random complete graph instances for two levels of sparsity.
$N$ and $M$ indicate the number of nodes and average number of edges in each of the $100$ instances.
\emph{\# Cor.} is the number of instances (out of $100$) in which BP-Bootstrap-Contract found the weight of the
optimal matching. \emph{\# LPs} indicates the average number of LPs solved by BP-Bootstrap. The bracketed numbers
indicate the minimum and maximum number of LPs solved across the $100$ instances. \emph{\# Blos.}
is the average number of blossoms collapsed in each MWPM problem.\label{table:complete_sparse}}
\end{center}
\end{table}

\begin{table}
\begin{center}
\begin{tabularx}{0.48\textwidth}{|l|l||c|c|c|} \hline
\footnotesize{N} & \footnotesize{M} & \scriptsize{\# Cor.} & \scriptsize{\# LPs} & \scriptsize{\# Blos.} \\ \hline
\footnotesize{100} & \footnotesize{285} & \footnotesize{95} & \scriptsize{5.7[2,13]} & \scriptsize{16.0[4,32]} \\ \hline
\footnotesize{250} & \footnotesize{733} & \footnotesize{83} & \scriptsize{10.6[4,28]} & \scriptsize{45.8[18,82]} \\ \hline
\footnotesize{500} & \footnotesize{1481} & \footnotesize{75} & \scriptsize{16.4[5,36]} & \scriptsize{92.1[44,132]} \\ \hline
\footnotesize{1000} & \footnotesize{2979} & \footnotesize{60} & \scriptsize{28.8[9,70]} & \scriptsize{200.3[140,294]} \\ \hline
\footnotesize{5000} & \footnotesize{14975} & \footnotesize{5} & \scriptsize{78.5[24,202]} & \scriptsize{996.0[796,1212]} \\ \hline
\end{tabularx}
\caption{Evaluation of BP-Bootstrap-Contract on random Triangulation instances.\label{table:triangulation}}
\end{center}
\end{table}

\begin{table}
\begin{center}
\begin{tabularx}{0.795\textwidth}{|l||c|c|c||l||c|c|c|} \hline
\multicolumn{4}{|c|}{Singly-Connected} & \multicolumn{4}{c|}{$3$-Connected} \\
\footnotesize{K} & \scriptsize{\# Cor.} & \scriptsize{\# LPs} & \scriptsize{\# Blos.} &
\footnotesize{K} & \scriptsize{\# Cor.} & \scriptsize{\# LPs} & \scriptsize{\# Blos.} \\ \hline

\scriptsize{100} & \scriptsize{100} & \scriptsize{10.0[3,17]} & \scriptsize{59.2[36,72]} &
\scriptsize{100} & \scriptsize{100} & \scriptsize{19.6[4,39]} & \scriptsize{60.3[18,102]} \\ \hline

\scriptsize{200} & \scriptsize{100} & \scriptsize{16.4[5,29]} & \scriptsize{128.1[86,156]} &
\scriptsize{200} & \scriptsize{100} & \scriptsize{34.3[5,71]} & \scriptsize{130.2[42,192]} \\ \hline

\scriptsize{500} & \scriptsize{100} & \scriptsize{34.4[11,68]} & \scriptsize{336.6[262,390]} &
\scriptsize{500} & \scriptsize{100} & \scriptsize{63.2[19,143]} & \scriptsize{355.8[200,466]} \\ \hline

\scriptsize{1000} & \scriptsize{100} & \scriptsize{58.8[19,111]} & \scriptsize{704.8[582,802]} &
\scriptsize{1000} & \scriptsize{100} & \scriptsize{119.7[20,223]} & \scriptsize{793.8[490,956]} \\ \hline

\end{tabularx}
\caption{Evaluation of BP-Bootstrap on a chain of triangles connected by a single edge or $3$ edges. \label{table:chained_triangles}}
\end{center}
\end{table}

We conducted a set of experiments to evaluate the performance of the BP-Bootstrap-Contract procedure. The BP-Bootstrap algorithm solves a series of LP problems, $LP^{(0)},...,LP^{(S)}$, where $LP^{(S)}$ is the first $LP$ with an integral solution. $LP^{(t+1)}$, the $LP$ in iteration $t+1$, is constructed from $LP^{(t)}$ by formulating a new weighted PM problem.

Since the $LP$ relaxation to the PM problem is half-integral~\cite{03Sch}, we are assured that the set of edges for which $\beta^{(t)}_{ij} = 1/2$ will form non-intersecting cycles of odd length (note that $\beta^{(t)}_{ij}$ denotes the belief on edge $(i,j)$ in the solution to $LP^{(t)}$). In addition, each odd-sized half-integral cycle identifies a constraint that can be used to tighten the relaxation of $LP^{(t)}$. Such constraints are referred to as \emph{blossom} constraints in the matching literature~\cite{Edmonds1965} and we will continue to adopt that name herein. Contracting all vertices correspondent to an odd-sized half-integral cycle has the effect of adding the blossom constraint to the $LP$. Thus, by contracting blossoms in each iteration, we hope to construct a series $LP^{(0)},...,LP^{(S)}$ of increasingly tight $LP$ relaxations to the PM problem.

Unfortunately, the procedure of greedily collapsing blossoms is not guaranteed to produce a tight $LP$ for the MWPM problem. This means that the BP-Bootstrap procedure will not necessarily recover the weight $E$ of the optimal perfect matching. We therefore rely on experimentation to gauge the efficacy of the proposed BP-Bootstrap heuristic.

We synthetically generated several types of weighted graph problems and compared the weight of the matching found by BP-Bootstrap to the weight of the optimal matching. The weight of the optimal perfect matching was found using Kolmogorov's BlossomV algorithm - the state-of-the-art algorithm for solving weighted matching problems~\cite{Kolmogorov2009}.

We conducted experiments on three problem types: 1) Sparse complete graph instances; 2) Triangulation instances; and 3) Chained triangle instance. The sparse complete graph instances were generated by taking a complete graph on $N$ nodes and eliminating edges with probability $p = \{0.5,0.9\}$.~\footnote{In doing so, we ensure that the graph contains a perfect matching by ensuring that the graph remains edge-connected} Edge weights were assigned an integral weight drawn from a discrete uniform distribution in $[1,2^{20}]$. The triangulation instances were generated by randomly placing points in the $2^{20}\times2^{20}$ square and computing a Delaunay triangulation on this set of points. The edge weights were set to the rounded Euclidean distance between two points. The chained triangle instances were used in the First DIMACS Implementation Challenge \cite{first-DIMACS-challenge}. Based on the algorithms t.f and tt.f by N. Ritchey and B. Mattingly, we generate a sequence of $K$ triangles. The chain of triangles is formed by connecting neighboring triangles via a single edge or by three edges. While certainly not exhaustive, this set of problems encompass an interesting range of graph types.

The results from running BP-Bootstrap on sparse complete graph instances are shown in Table \ref{table:complete_sparse}. For each number of nodes $N \in \{100,250,500,1000 \}$, we generated $100$ random instances. We report the average number of edges, $M$, for each setting of $N$ and \emph{\# Cor.}, the number of instances (out of $100$) for which BP-Bootstrap found the optimal weight matching. In addition, we report the average and [minimum, maximum] number of $LPs$ solved by BP-Bootstrap as well as the number of blossoms contracted by the BP-Bootstrap procedure. From this table it seems clear that BP-Bootstrap procedure is accurate on dense complete graphs. A reason for the success of the method on this class of graphs is that $LP^{(0)}$, the initial $LP$ relaxation, is tight a large percentage of the time. This is indicated by the fact that on average less than two $LPs$ need to be solved by BP-Bootstrap.

In contrast, the Bootstrap method performs rather poorly on the random Triangulation instances shown in Table \ref{table:triangulation}. The greedy blossom contracting approach does not appear to be successful on this class of problems. In fact, when we examined the operation of Kolmogorov's BlossomV algorithm on these instances we found that BlossomV frequently needed to backtrack and expand blossoms that had been collapsed - an operation not supported by our greedy heuristic.

Finally, the results from running BP-Bootstrap on the chained triangle instances are shown in Table \ref{table:chained_triangles}. The BP-Bootstrap method performs well on these problems. Interestingly, we see that the greedy heuristic is effective even though these instances require solving many $LPs$ and collapsing many blossoms.

\section{Questions yet to be addressed in this project}
\label{sec:path}

We conclude by describing what we believe are the two most important challenges needing to be resolved in our ``Matching" using BP project.

\begin{itemize}
\item As proven in \cite{11Gur} for the bipartite (permanent) case: $Z\geq Z_{BP}$. As noticed in \cite{11CY}, the proof of \cite{11Gur} is in fact a direct corollary of Eq.~(\ref{tilde_beta}) combined with a combinatorial result of Schrijver from \cite{98Sch}.  Indeed, it was proven in \cite{98Sch} (see Corollary 1c) that $\mbox{Perm}(\beta.(1-\beta))\geq \prod_{(i,j)\in {\cal E}}(1-\beta_{ij})$,  which is essentially the statement that the rhs of Eq.~(\ref{tilde_beta}) in the bipartite case is equal or larger than unity. The proof of Schrijver is indirect and the author mentions in passing ``... we have tried to find a direct proof of it, based on continuity and differentiability, and did not succeed".  Is it possible to find such a direct proof based on the combination of the Loop Calculus (\ref{Loop},\ref{rC}) and the Bootstrap-BP (\ref{tilde_beta}) formulas in the bipartite case? Can one generalize the $Z\geq Z_{BP}$ statement and Schrijver inequality to the case of a general, non-bipartite, graph?

\item In general, and as explained above, our Bootstrap-and-Contract algorithm outputs an upper bound on the MWPM. In our empirical study, we observed that when our algorithm fails to produce an optimal matching, the BlossomV algorithm needs to backtrack and expand one or more contracted blossoms. A remaining question is how to incorporate the undo/expand step into our procedure. In particular, can one do so in a distributed manner, via a message passing (or perhaps just greedy) procedure?

\end{itemize}

\appendix

\section{Convexity of the Bethe Free Energy}
\label{app:conv}

This Appendix generalizes the main result of \cite{13Von}, where convexity of the BFE was proven for the case of PM GM over the bi-partite graph, to the case of PM GM over a general graph. Indeed, according to Theorem 20 of \cite{13Von}, $\sum_{j:(i,j)\in {\cal E}}\log \beta_{ij}-(1-\beta_{ij})\log(1-\beta_{ij})$ is convex within the polytope defined by, $\sum_{j:(i,j)\in {\cal E}} \beta_{ij}=1$ and $\beta_{ij}\geq 0$, for all $(i,j)\in {\cal E}$ and regardless of the global structure of the graph ${\cal G}$. Therefore, ${\cal F}_{\mbox{BP}}$ (\ref{BFE}) is convex within the domain $\mbox{DS}$ simply because it is represented as a sum of functions which are all convex within the domain.

\section{Derivation of the bootstrap formulas}
\label{app:boot}

In this Appendix  we derive Eqs.~(\ref{tilde_beta},\ref{tilde_beta_gamma}) following the logic of \cite{10WC},  see also Appendixes B.1, D.1 of \cite{11CY}. For the purpose of generality we will consider arbitrary (fractional) $\gamma$ and arbitrary graph ${\cal G}$.

Our algebraic derivation consists of comparing two formulas. First, considering vectors of weights corresponding to the two sides of Eq.~(\ref{BP-gamma}) and equating the respective partition functions one derives
\begin{eqnarray}
Z(\tilde{\beta}_\gamma)=\sum_{\sigma\in\Sigma}
\exp\left(\sum_{(i,j)\in{\cal E}}\sigma_{ij}\frac{\mu_i+\mu_j-\varepsilon_{ij}}{T}\right)=
Z(w)\exp\left(\frac{\sum_{i\in {\cal V}}\mu_i}{T}\right),
\label{Boot1}
\end{eqnarray}
where we have used, $\sum_{(i,j)\in{\cal E}}\sigma_{ij}(\mu_i+\mu_j)=\sum_{i\in{\cal V}}\mu_i
\sum_{j:(i,j)\in{\cal E}}\sigma_{ij}=\sum_{i\in{\cal V}}\mu_i$. Second, let us apply $\log$ to both sides of Eqs.~(\ref{BP-gamma}) and then sum the results over all the edges of ${\cal G}$ weighted with $\beta_{ij}$
\begin{eqnarray}
\sum_{(i,j)\in{\cal E}}\beta_{ij}\left(\log\beta_{ij}-\gamma\log(1-\beta_{ij})\right)=\frac{1}{T}\left(\sum_{i\in{\cal V}}\mu_i -\sum_{(i,j)\in{\cal E}}\sigma_{ij}\varepsilon_{ij}\right),
\label{Boot2}
\end{eqnarray}
where we have used the double stochasticity of the optimal $\beta$ through, $\sum_{(i,j)\in{\cal E}}\beta_{ij}(\mu_i+\mu_j)=\sum_{i\in{\cal V}}\mu_i
\sum_{j:(i,j)\in{\cal E}}\beta_{ij}=\sum_{i\in{\cal V}}\mu_i$. Then, combining Eqs.~(\ref{Boot1},\ref{Boot2}) with Eq.~(\ref{Zgamma}) we arrive at Eq.~(\ref{tilde_beta_gamma}). $\blacksquare$

\section*{References}

\bibliographystyle{iopart-num}%{abbrv}%apsrev4-1
\bibliography{Bib/permanent,Bib/BP_review,Bib/zeta,Bib/MishaPapers,Bib/refs_matching}

\end{document}